\documentclass[fleqn,11pt]{article}
\usepackage{amsmath,amssymb,amsfonts,graphicx}

\usepackage{amsfonts,amssymb,cite}

\newcommand{\bls}[1]{\renewcommand{\baselinestretch}{#1}}

\def\noi{\noindent}

\newcommand{\Title}[1]{\noi {{\Large\bf #1}}\\[1ex]}

\def\Aunames#1{\noi{\bf #1}}
\def\au#1{${}^{#1}$}
\def\Addresses#1{\medskip\noi \protect
	\begin{description}\itemsep -3pt {\it #1} \end{description}}
\def\adr#1#2{\item[${}^{#1}$]{\it #2}}

\newcommand{\Abstract}[1]{\vskip 2mm \begin{center}
        \parbox{16.4cm}{\small\noi #1} \end{center}\medskip}

\def\email#1#2{\footnotetext[#1]{e-mail: #2}\addtocounter{footnote}{1}}

\def\nqq{\hspace*{-2em}}
\def\nhq{\hspace*{-0.5em}}

\def\cm{\hspace*{1cm}}
\def\inch{\hspace*{1in}}

\def\Jl#1#2{#1 {\bf #2},\ }

\def\ApJ#1 {\Jl{Astroph. J.}{#1}}
\def\CQG#1 {\Jl{Class. Quantum Grav.}{#1}}
\def\DAN#1 {\Jl{Dokl. AN SSSR}{#1}}
\def\GC#1 {\Jl{Grav. Cosmol.}{#1}}
\def\GRG#1 {\Jl{Gen. Rel. Grav.}{#1}}
\def\JETF#1 {\Jl{Zh. Eksp. Teor. Fiz.}{#1}}
\def\JETP#1 {\Jl{Sov. Phys. JETP}{#1}}
\def\JHEP#1 {\Jl{JHEP}{#1}}
\def\JMP#1 {\Jl{J. Math. Phys.}{#1}}
\def\NPB#1 {\Jl{Nucl. Phys. B}{#1}}
\def\NP#1 {\Jl{Nucl. Phys.}{#1}}
\def\PLA#1 {\Jl{Phys. Lett. A}{#1}}
\def\PLB#1 {\Jl{Phys. Lett. B}{#1}}
\def\PRD#1 {\Jl{Phys. Rev. D}{#1}}
\def\PRL#1 {\Jl{Phys. Rev. Lett.}{#1}}

\def\al{&\nhq}
\def\lal{&&\nqq {}}
\def\eq{Eq.\,}
\def\eqs{Eqs.\,}
\def\beq{\begin{equation}}
\def\eeq{\end{equation}}
\def\bear{\begin{eqnarray}}
\def\bearr{\begin{eqnarray} \lal}
\def\ear{\end{eqnarray}}
\def\earn{\nonumber \end{eqnarray}}
\def\nn{\nonumber\\ {}}
\def\nnv{\nonumber\\[5pt] {}}
\def\nnn{\nonumber\\ \lal }

\def\yyy{\\[5pt] \lal }
\def\eql{\al =\al}

\def\tst{\textstyle}

\def\fract#1#2{{\tst\frac{#1}{#2}}}

\def\half{{\fract{1}{2}}}

\def\e{{\,\rm e}}
\def\d{\partial}

\def\sign{\mathop{\rm sign}\nolimits}

\def\const{{\rm const}}

\def\then{\ \Rightarrow\ }

\def\eqn#1{\eq\eqref{#1}}
\def\rf{\eqref}

\def\mn{_{\mu\nu}}
\def\MN{^{\mu\nu}}
\def\mN{_\mu^\nu}

\def\ax{_{\rm ax}}
\def\thd{\tfrac 13}

\def\cR{{\cal R}}

\def\kappa{\varkappa}

\def\hG{{\hat\Gamma}}

\def\R{{\mathbb R}}

\def\og{{\bar g}}
\def\ophi{{\bar\phi}}
\def\oR{{\bar R}}

\def\ME{\mbox{$\mathbb{M}_{\rm E}$}}
\def\MJ{\mbox{$\mathbb{M}_{\rm J}$}}
\def\GR{general relativity}
\def\sph{spherically symmetric}
\def\ssph{static, spherically symmetric}
\def\cy{cylindrical}
\def\cyl{cylindrically symmetric}

\def\bh{black hole}
\def\bhs{black holes}
\def\wh{wormhole}
\def\whs{wormholes}
\def\asflat{asymptotically flat} 

\usepackage{color}

\bls{1.01}
\begin{document}
\thispagestyle{empty}
\twocolumn[

\bigskip

\Title {Hybrid metric-Palatini gravity: Regular stringlike configurations}
	
\Aunames{K. A. Bronnikov,\au{a,b,c,1} S. V. Bolokhov,\au{b;2} and M. V. Skvortsova\au{b;3}}

\Addresses{\small
\adr a {Center of Gravitation and Fundamental Metrology, VNIIMS, 
		Ozyornaya ul. 46, Moscow 119361, Russia}
\adr b {Institute of Gravitation and Cosmology, Peoples' Friendship University of Russia
		(RUDN University), \\ ul. Miklukho-Maklaya 6, Moscow 117198, Russia}
\adr c{National Research Nuclear University ``MEPhI'', 
		Kashirskoe sh. 31, Moscow 115409, Russia}
        }
                
\Abstract
   {We discuss static, cylindrically symmetric vacuum solutions of hybrid metric-Palatini 
    gravity (HMPG), a recently proposed theory that has been shown to successfully 
    pass the local observational tests and to produce a certain progress in cosmology.
    We use HMPG in its well-known scalar-tensor representation. The latter coincides with 
    general relativity containing, as a source of gravity, a conformally coupled scalar field $\phi$ 
    and a self-interaction potential $V(\phi)$. The $\phi$ field can be canonical or phantom,
    and accordingly the theory splits into canonical and phantom sectors. 
    We seek solitonic (stringlike) vacuum solutions of HMPG, that is, completely regular
    solutions with Minkowski metric far from the symmetry axis, with a possible angular 
    deficit. A transition of the theory to the Einstein conformal frame is used as a tool, and 
    many of the results apply to the general Bergmann-Wagoner-Nordtvedt class of scalar-tensor 
    theories as well as $f(R)$ theories of gravity. One of these results is a one-to-one 
    correspondence between stringlike solution in the Einstein and Jordan frames if the conformal 
    factor that connects them is everywhere regular. An algorithm for construction of stringlike 
    solutions in HMPG and scalar-tensor theories is suggested, and some examples
    of such solutions are obtained and discussed.
    }

\bigskip

] 
\email 1 {kb20@yandex.ru} 
\email 2 {boloh@rambler.ru}
\email 3 {milenas577@mail.ru}

\section{Introduction}

  General relativity (GR) is well known to be quite successful in describing local observational 
  effects in the Solar system in stellar astrophysics and, quite probably, in \bh\ physics. However, 
  it faces serious problems at larger scales: on the galactic scale, it does not give a satisfactory 
  explanation of the rotation curves without introducing the so-called Dark Matter (DM) of still 
  unknown nature, and it cannot account for the observed accelerated expansion of the Universe
  without introducing the so-called Dark Energy (DE), an unknown kind of matter with large negative 
  pressure.  According to the latest observations,
  the energy content of the Universe consists, in terms of GR, of only about 5\,\% of visible
  matter, about 20\,\% of DM and the remaining 75\,\% of DE, see, e.g., \cite{ishak-rev,planck}. 

  In addition to attempts to solve the DM and DE problems in the framework of GR by 
  introducing so far unobservable forms of DM like WIMPs (weakly interacting massive
  particles) and various forms of DE like cosmological constant or ``quintessence''
  scalar fields, etc., \cite{DE}, there is an alternative broad trend, addressing the same 
  problems using various extensions of GR, such as, for instance, $f(R)$ theories, theories
  with additional scalar, vector and spinor fields, multidimensional theories and those with
  extensions of the Riemannian geometry (metric-affine theories, those with Finslerian 
  geometry, etc. \cite{ex-GR1, ex-GR2}).

  One of such recently proposed extensions of GR is the Hybrid Metric-Palatini Gravity (HMPG) 
  \cite{har12}. In this theory, one assumes the independent existence of the Riemannian metric 
  $g\mn$ and the connection $\hG\mn^\alpha$. The total action of HMPG reads \cite{har12}
\beq  				\label{S}
	S = \frac{1}{2\kappa^2}\int d^4 x\sqrt{-g} [R + F(\cR)] + S_m,
\eeq    
  where $R = R[g]$ is the Ricci scalar derived as usual from $g\mn$, while $F(\cR)$ is an arbitrary
  function of the scalar $\cR = g\MN \cR\mn$ obtained with the Ricci tensor $\cR\mn$ calculated
  in the standard way from the connection $\hG\mn^\alpha$; furthermore, $g = \det(g\mn)$,
  $\kappa^2$ is the gravitational constant, and $S_m$ is the action of all nongravitational matter.
  
  HMPG, which combines the metric and Palatini approaches to the description of gravity and 
  extends the formulation of $f(R)$ theories, has a number of achievements described in the 
  reviews  \cite{cap15-rev, har18-book, har20-rev}. In particular, it agrees with the classical
  tests in the Solar system \cite{cap13b}, fairly well solves the DM problem concerning the
  dynamics of galaxies and galaxy clusters; it has been shown to be able to describe an
  accelerating Universe without invoking a cosmological constant \cite{cap13}.
  Some papers have been devoted to the possible existence and properties of vacuum \ssph\
  \bhs\ and \whs\ in HMPD \cite{dan19, kb19, we20} as well as static \cyl\ configurations 
  intended to reproduce the basic properties of cosmic strings \cite{har20-cy}.

  Let us also mention a further generalization of HMPG, containing an arbitrary function of 
  the two curvature scalars $R$ and $\cR$, and developed in  
  \cite{boh13, mont19, lem19, lem20}.
  
  The present paper is devoted to a study of vacuum static \cyl\ solutions of HMPG. The metric is 
  assumed in the general form 
\beq                                                      		   \label{ds}
	    ds^2 = \e^{2\gamma} dt^2 - \e^{2\alpha} dx^2  -\e^{2\mu} dz^2 - \e^{2\beta} d\varphi^2
\eeq
   where $\alpha,\beta,\gamma,\mu$ are functions of $x$; $z\in\R$ and $\varphi \in [0,2\pi)$ 
   are the longitudinal and azimuthal coordinates, respectively, and the radial coordinate $x$
   is arbitrary and admits reparametrization to any smooth monotonic function of $x$. Our 
   goal will be to find regular stringlike configurations, that is, HMPG solutions with the metric 
   \rf{ds} having a regular axis and a proper behavior at large values of the circular radius 
   $r \equiv \e^\beta$: the latter means that the metric should be either \asflat\ or flat up to 
   an angular deficit, which will then be proportional to the effective cosmic string tension.
   Under these conditions, such a globally regular solitonlike field configuration can be observed 
   from a distant flat or very weakly curved region of space like a cosmic string.   

  The same objective was formulated in \cite{har20-cy}, where the authors used the 
  coordinate condition $\alpha = \gamma$ in our notations (that is, $g_{tt} = - g_{xx}$, see \eq (17)
  in \cite{har20-cy}, and note that there the radial coordinate is denoted $r$ instead of our $x$).
  In addition, based on the desired string interpretation, it was postulated there that the metric 
  should be invariant under boosts in the $z$ direction \cite{vilenkin81},
  which leads to the requirement $g_{tt} = -g_{zz}$. As a result, solutions were sought for 
  in \cite{har20-cy} under the restrictive condition $\alpha=\gamma=\mu=0$
  in the present notations, with only one unknown metric function $\e^{\beta(x)}$ (equal to $W(r)$ 
  in \cite{har20-cy}, see \eq (29) there). A number of analytical and numerical solutions were 
  then obtained and discussed, but none of them possessed a regular axis and therefore 
  none of them were able to represent a regular extended cosmic string model. 

  In this paper we undertake an extended study of the same problem. First of all, we do not 
  impose any restriction on the metric \rf{ds} (except for a convenient choice of the radial coordinate $x$)
  but show that the requirement of a regular axis and a regular (that is, flat or string) asymptotic 
  behavior inevitably implies $g_{tt} = -g_{zz}$, which leads to boost invariance in the $(t,z)$ subspace
  and makes it unnecessary to separately require this invariance. Furthermore, we here restrict ourselves 
  to seeking only stringlike solutions, formulate some necessary conditions for their existence 
  and present a few examples of such solutions. (We will use the terms ``stringlike'' and ``solitonic'' 
  as synonyms.)

  As in \cite{har20-cy}, we will also employ the scalar-tensor theory (STT) representation of HMPG, but,
  unlike these authors, we essentially use the well-known conformal mapping leading to the Einstein frame,
  which formally coincides with GR with a minimally coupled scalar field as the source of gravity. The 
  latter problem has been studied in \cite{kb01-cy}, and a number of results and observations obtained 
  there turn out to be useful for studying the present problem. More general data on the necessary 
  mathematical definitons concerning \cy\ space-times and discussions of numerous \cyl\ solutions 
  obtained in GR with different sources of gravity can be found in the book \cite{exact-book} and the 
  review \cite{BSW19}.  
       
  The paper is organized as follows. In Section 2 we briefly recall the main features of HMPG and its
  STT representation \cite{har12,cap15-rev}. In Section 3 we formulate the problem to be solved, including 
  the explicit form of the field equations and the boundary conditions of regularity at large and small 
  radii. In Section 4 we reproduce the results of \cite{kb01-cy} relevant to our present problem and 
  use them to obtain some necessary conditions for the existence of stringlike solitonic solutions in HMPG.
  Section 5 is devoted to particular examples of such solutions, and Section 6 contains a discussion and   
  some concluding remarks.
  
\section {HMPG and its scalar-tensor representation}  
  
  As is known from \cite{har12,cap15-rev,dan19}, varying the action \rf{S} with respect to the connection 
  $\hG\mn^\alpha$, we obtain that this connection coincides with the familiar Levi-Civita connection
  for the metric $h\mn = \phi g\mn$ conformal to $g\mn$, where the conformal factor is
  $\phi = F_\cR \equiv dF/d\cR$. It clearly shows that the HMPG actually involves,
  in addition to $g\mn$, a single additional dynamic degree of freedom expressible in the
  scalar field $\phi$. As is demonstrated in \cite{har12,cap15-rev}, the whole theory can 
  be reformulated as a specific example of a scalar-tensor theory (STT) where the gravitational 
  part of the action is
\beq                           \label{S1} 
	S_g = \! \int \! d^4x \sqrt{-g}\bigg[(1+\phi)R - \frac {3}{2\phi}(\d\phi)^2 - V(\phi)\bigg],
\eeq          
   where\footnote
	  {Unlike the papers \cite{har12, cap15-rev, dan19} etc., we use the metric signature 
	  $(+ - -\, -)$, therefore, the plus before $(\d\phi)^2 = g\MN \phi_{\mu}\phi_{\nu}$ 
	  refers to a canonical field, and a minus means that the field is phantom. Also,
	  in what follows we consider only vacuum HMPG solutions ($S_m =0$) and safely omit 
	  the factor $1/(2\kappa^2)$ near the gravitational part of the action. The Ricci tensor 
	  is defined as $R\mn =\d_\nu \Gamma^\alpha_{\mu\alpha} - \ldots$, thus, for example, 
	  the Ricci scalar $R$ is positive for de Sitter space-time. The units are used in which 
	  the speed of light and the Newtonian gravitational constant are equal to unity. } 
   the potential $V(\phi) $ is expressed in terms of $f(\cR)$:
\beq                             \label{VR}
		V(\phi) = \cR F_\cR - F(\cR).
\eeq    
 
  The theory \rf{S1} belongs to the Bergmann-Wagoner-Nordtvedt class of STT 
   \cite{berg68, wag70, nor70}, characterized by the gravitational action 
\beq   	 	\label{S-STT}
		S_g =\int\! d^4x \sqrt{-g}\Big[f(\phi)R  + h(\phi)(\d\phi)^2 - V(\phi)\Big],
\eeq     
  where $f(\phi)$, $h(\phi)$ and $V(\phi)$ are arbitrary functions. In the present case, $V(\phi)$
  is given by \rf{VR}, and
\beq  			\label{f,h-I}
		f(\phi) = 1+\phi, \cm     h(\phi) = - \frac{3}{2\phi}. 	
\eeq 
    
  The general action \rf{S-STT} is known to admit a conformal mapping  \cite{wag70}
  to the Einstein frame \ME\ in which the scalar field is minimally coupled to the metric 
  (the formulation \rf{S-STT} is said to correspond to the Jordan conformal frame 
   \MJ). The transformation is given by \cite{wag70}
\bearr 			\label{J-E}
	\og\mn = f(\phi) g\mn, \qquad \frac {d\phi}{d\ophi} = f (\phi) |D(\phi)|^{-1/2},
\nnn	
	D(\phi) = f(\phi)h(\phi) + \frac 32 \bigg(\frac{df}{d\phi}\bigg)^2,
\ear
  and leads to the action in \ME
\beq  	\nhq		\label{S-E}
	S_g = \int \! d^4x \sqrt{-\og}
	\bigg[\oR + n \og\MN \ophi_{,\mu}\ophi_{,\nu} - \frac{V(\phi)}{f^2(\phi)}\bigg],
\eeq                      
  where quantities obtained from or with the transformed metric $\og\mn$ are marked by
  overbars. The factor $n = \sign D(\phi)$ distinguishes two kinds of scalar fields: $n=+1$ 
  corresponds to canonical scalars with positive kinetic energy, while $n =-1$ describes 
  phantom fields with negative kinetic energy.
  
  In the theory \rf{S1} we obtain $D = -3/(2\phi)$, so that $n = - \sign \phi$, and, as a result,
\bear      \nhq \label{can}
	\phi = -\tanh^2 \frac{\ophi}{\sqrt 6} && (n = +1,\ \  -1 < \phi <0)  ,
\\                  \label{pha}
	\phi = \tan^2 \frac{\ophi}{\sqrt 6} \quad\  &&   (n = -1, \ \  \phi > 0). 
\ear    
  Thus, according to the sign of $\phi$, the whole theory splits into two sectors, the canonical 
  one and the phantom one. We also notice that values of $\phi$ smaller than $-1$ are apparently
  physically meaningless because they lead to a negative effective gravitational constant.
  
  Substituting $\phi = - n \chi^2/6$ in the action \rf{S1}, we convert it to
\bearr \label{S2}
		S_g = \int \! d^4x \sqrt{-g} \Big[(1 - n \chi^2/6)R 
\nnn \inch	\cm	
		+ n (\d\chi)^2 - W(\chi)\Big],
\ear
  where $W(\chi) = V(\phi)$. This action corresponds to GR where the only source of
  gravity is a conformally coupled scalar field whose kinetic energy has the usual sign 
  if $\phi < 0$ ($n =1$) and the anomalous sign if $\phi > 0$ ($n = -1$).
  Conformally coupled scalar fields have been studied by many authors, beginning with Penrose 
  \cite{pen64} (who considered a massless conformally invariant scalar field) and Chernikov and 
  Tagirov \cite{tag68} (who introduced and analyzed massive conformally coupled fields). 
  The phantom version of \rf{S2} was discussed in \cite{ZK} as a possible alternative to GR 
  in different cosmological and astrophysical applications. 
   
  The transition \rf{J-E} has been quite often used for finding exact or approximate solutions
  to the STT field equations because the equations due to \rf{S-E} are appreciably
  simpler than those due to the action \rf{S-STT}. Having found an Einstein-frame solution,
  it is easy to obtain its Jordan-frame counterpart using the transformation inverse to \rf{J-E}.
  
  However, one should bear in mind an important subtle point: if the function $f(\phi)$ 
  in \rf{S-STT} is singular (zero or infinity) at some value of $\phi$, it may happen that 
  a singularity in \ME\ with the metric $\og\mn$ maps to a regular surface in \MJ\
  with the metric $g\mn$ (or vice versa), and then the manifold \MJ\ is continued
  beyond such a surface. This phenomenon has been termed {\it conformal continuation}
  \cite{kb01p,kb02}, and it takes place in many scalar-vacuum and scalar-electrovacuum
  solutions of STT and $f(R)$ theories of gravity \cite{kb02, kb-ch05}, including, in particular,
  \sph\ solutions of GR with a conformal scalar field \cite{kb70,kb73} and those
  of the Brans-Dicke theory \cite{CBH1, CBH2}. In the present paper we will meet 
  conformal continuations while dealing with \cyl\ manifolds. 

  Using the STT representation, it is natural to ask: if we know a solution with a certain 
  potential $V(\phi)$, what is then the original form of HMPG, on other words, 
  what is the corresponding function $F(\cR)$? The answer follows from \eqn {VR}. In the case 
  $V(\phi) \equiv 0$ we have simply $F(\cR) = \const\cdot \cR$.
  In the general case $V(\phi) \not\equiv 0$, since $\phi = F_\cR$, the relation \rf{VR} has the form
  of a Clairaut equation with respect to $F(\cR)$ (see, e.g., \cite{kamke}); its solution consists 
  of a regular family containing only linear functions,
\beq         \label{cle1}
		F(\cR) = H\cR - V(H), \qquad H = \const,
\eeq      
  and the so-called singular solution, which forms an envelope of the regular family of solutions 
  and can be written in a parametric form:
\bearr                       \label{cle2}
		F(\cR) = \phi\cR - V(\phi),
\nnn
		\cR = dV/d\phi.
\ear            
   A more detailed discussion of this issue can be found in \cite{dan19}.
     
\section{Cylindrical symmetry: Equations and boundary conditions}

\def\aJ{\alpha_{_{\rm J}}}
\def\bJ{\beta_{_{\rm J}}}
\def\gJ{\gamma_{_{\rm J}}}
\def\mJ{\mu_{_{\rm J}}}
\def\rJ{r_{_{\rm J}}}
\subsection{Equations}

  The final results for HMPG must be formulated in the Jordan frame \MJ, 
  for which we assume the metric \rf{ds} in slightly different notations,
\beq    \nhq                                                  		   \label{ds_J}
	    ds_J^2 = \e^{2\gJ} dt^2 \!-\! \e^{2\aJ} dx^2 \! -\! \e^{2\mJ} dz^2 \! - \! \e^{2\bJ} d\varphi^2,
\eeq
  where $\aJ, \bJ, \gJ, \mJ$ are functions of $x$, as well as the scalar field $\phi$.
  However, we will seek solutions to the field equations in the Einstein frame \ME,
  corresponding to the action \rf{S-E}, in which, for convenience, we make the 
  substitution  $\ophi \to \psi$:
\bearr  	\nhq		\label{S-E1}
	S_g = \int \! d^4x \sqrt{-\og}\Big[\oR + 6n (\bar{\d}\psi)^2 - U(\psi)\Big],
\nnn
		\psi = \ophi/\sqrt{6}, \cm (\bar{\d}\psi)^2 = \og\MN \psi_{,\mu}\psi_{,\nu},
\nnn
		U(\psi) = \frac {V(\phi)}{(1+\phi)^2}.
\ear
   The corresponding field equations read
\bearr     \label{e-psi}
	   12 n \bar{\Box} \psi + dU/d\psi =0,	 
\yyy              \label{EE}
	   \oR\mN - \half \delta\mN \oR = - T\mN [\psi],
\ear
  where $\bar{\Box}$ is the d'Alambertian operator defined for the metric $\og\mn$, and
\beq                \label{SET}
		T\mN[\psi] = 6n\Big[\psi_{,\mu}\psi^{,\nu} - \half\delta\mN (\bar{\d}\psi)^2\Big] 
					+ \half\delta\mN U(\psi).
\eeq
   These field equations will be considered for the scalar $\psi = \psi (x)$ and the metric 
\beq                                                      		   \label{ds_E}
	    ds_E^2 = \e^{2\gamma} dt^2 - \e^{2\alpha} dx^2  -\e^{2\mu} dz^2 - \e^{2\beta} d\varphi^2.
\eeq
   The Einstein- and Jordan-frame quantities are related by
\bearr                               \label{cosh}
       \phi = - \tanh^2 \psi, \qquad ds_J^2 = \cosh^2\psi \ ds_E^2
\nnn
			\inch (n =1,\ \mbox{canonical sector}),
\yyy                                  \label{cos}
       \phi = \tan^2 \psi, \qquad ds_J^2 = \cos^2\psi \ ds_E^2
\nnn
			\inch (n = -1,\ \mbox{phantom sector}).
\ear
   
   It is convenient to use the alternative form of the Einstein equations
\beq                                      \label{EE'}
     \oR\mN = - 6n \psi_{,\mu}\psi^{,\nu} +  \half\delta\mN U(\psi);
\eeq
  the nonzero components of the Ricci tensor $\oR\mN$ are
\bear                 \label{Ric}
      \oR^0_0 \eql -\e^{-2\alpha}[\gamma'' + \gamma'(\sigma' -\alpha')] ,
\nnv      
      \oR^1_1 \eql -\e^{-2\alpha}[\sigma'' + \sigma'{}^2 
\nnn \cm
			- 2(\beta'\gamma'  + \beta'\mu' + \gamma' \mu') - \alpha'\sigma'],
\nnv      
      \oR^2_2 \eql -\e^{-2\alpha}[\mu'' + \mu'(\sigma' -\alpha')],  
\nnv      
      \oR^3_3 \eql -\e^{-2\alpha}[\beta'' + \beta'(\sigma' -\alpha')], 
\ear
  where the prime denotes $d/dx$, and we have introduced the notation
\beq                                    
            \sigma = \beta + \gamma + \mu;
\eeq     
  it is also helpful to write the constraint equation from \rf{EE}, which is the first integral of the 
  others and contains only first-order derivatives of the metric functions:
\beq                  \label{G11}
	      \beta'\gamma'  + \beta'\mu' + \gamma' \mu' 
								= 3n \psi'{}^2 - \half U(\psi) \e^{2\alpha}.
\eeq

\subsection{Boundary conditions}       

  Let us, for convenience, formulate the boundary conditions in the Einstein frame, 
  referring to their more detailed description in \cite{kb01-cy}. Similar conditions for \MJ\ will 
  only require putting the index ``J'' near each of the letters $\alpha,\beta,\gamma,\mu$.

  A {\bf regular axis} means that there is some value $x = x\ax$ at which the circular radius 
  $r = \e^\beta \to 0$ while the algebraic curvature invariants remain finite. 
  Let us note that the Kretschmann invariant $K = R^{\mu\nu\rho\sigma}R_{\mu\nu\rho\sigma}$ 
  for the metric \rf{ds_E} is a sum of squared components of the nonzero Riemann tensor
  components $R\MN{}_{\rho\sigma}$:
\bear
    K \eql 4 \sum_{i=1}^{6} K_i^2; 
\nn
    K_1 \eql R^{01}{}_{01}
    		=-\e^{-\alpha-\gamma}(\gamma'\e^{\gamma-\alpha})',
\nn
    K_2 \eql R^{02}{}_{02} = -\e^{-2\alpha}\gamma'\mu',
\nn
    K_3 \eql R^{03}{}_{03} = -\e^{-2\alpha}\beta'\gamma',
\nn
    K_4 \eql R^{12}{}_{12} = -\e^{-\alpha-\mu}(\mu'\e^{\mu-\alpha})',
\nn
    K_5 \eql R^{13}{}_{13} = -\e^{-\alpha-\beta}(\beta'\e^{\beta-\alpha})',
\nn
    K_6 \eql R^{23}{}_{23} = -\e^{-2\alpha}\beta'\mu'              \label{Kr}
\ear
   Thus for $K < \infty$ it is necessary and sufficient that all $|K_i| <  \infty$, which in turn 
   guarantees finite values of all algebraic invariants composed from the Riemann tensor. 
   It is important that all $K_i$ in \rf{Kr} are independent of the choice of the coordinate $x$.

  One can verify \cite{kb01-cy} that all $K_i$ are finite on the axis $r = \e^\beta \to 0$ if and only if 
  $\gamma(x)$ and $\mu(x)$ tend to finite limits $\gamma\ax$ and $\mu\ax$, and
\bearr              \label{ax1}
         \gamma'\e^{-\alpha} = O(r),\qquad  \mu'\e^{-\alpha} = O(r),
\yyy                \label{ax2}
		|\beta'| \e^{\beta-\alpha}= 1 + O(r^2).  
\ear
   the latter condition expressing a correct relation between an infinitesimal circumference and 
   its radius, in other words, the absence of a conical singularity on the axis.\footnote
	  {We denote by $O(f)$ a quantity of either the same order of magnitude as $f$ or smaller 
	    in a certain limit, while quantities of the same order are connected by the symbol $\sim$.}
 
   A correct {\bf asymptotic behavior}, guaranteeing that our configuration will be visible for a 
  distant observer, requires a zero curvature limit, which in turn implies finite values of  
  $\gamma(x)$ and $\mu(x)$
  at some $x = x_{\infty}$ where $r = \e^\beta \to \infty$. Moreover, a flat-space limit 
  at spatial infinity would require a condition similar to \rf{ax2}, but a cosmic string geometry 
  admits a more general condition,
\beq                          \label{defect}
		|\beta'| \e^{\beta-\alpha}\to 1 - \mu_s, \cm \mu_s = \const < 1,
\eeq 
  where $2\pi \mu_s$ is the angular deficit characterizing the gravitational field of a cosmic string..
  Thus the space-time is locally flat but globally behaves as if there were a conical singularity.
  Following \cite{kb01-cy}, we will call such a flat or stringlike asymptotic behavior at large 
  $\e^\beta$ ``{\it a regular asymptotic}''.

  As can be easily verified (for example, by using the Gaussian normal radial coordinate specified
  by the condition $\e^\alpha \equiv 1$), under the above regularity conditions on the axis and at 
  infinity, the total energy of matter per unit length along the $z$ axis is finite. This quantity
  is determined by the integral 
\beq                                           \label{E_tot}
	 {\cal E} =  \int T^0_0 \e^{\alpha+\beta+\mu} dx\,dz\,d\varphi 
\eeq
  (where integration in $z$ covers a unit interval), which converges at infinity because the Ricci 
  tensor components \rf{Ric} decay there as $r^{-3}$ or faster (assuming that all quantities can 
  be expanded in power series in $1/r$), and the same is true for $T\mN$ due to the Einstein 
  equations. Recalling the expression \rf{SET} for $T\mN$, we conclude that  at a regular
  asymptotic both quantities $U$ and $\e^{-2\alpha} \psi'{}^2$ decay at infinity as $r^{-3}$ or 
  faster.
  
  The angular deficit $2\pi \mu_s$ is directly proportional to the quantity \rf{E_tot} in the simplest 
  string model with a flat metric everywhere except the symmetry axis, where a conical 
  singularity takes place \cite{vilenkin81}. Such a relationship is not evident in the general case.

  We are seeking solitonic solutions which possess both a regular axis and a regular asymptotic.

 \section{Stringlike solutions: Analysis}
\subsection{Solitons in the Einstein frame}

  In this subsection we briefly reproduce and discuss some results of \cite{kb01-cy}, 
  relevant to our study, and add some more observations.

  Let us choose the harmonic $x$ coordinate in the metric \rf{ds_E}, such that\footnote
   	{To our knowledge, this coordinate condition was used for the first time for 
	finding \cyl\ solutions in GR in \cite{kb79-cy}.}
\beq
    		\alpha = \beta + \gamma + \mu                             \label{harm}
\eeq
  (note that such a condition does not hold in the corresponding Jordan-frame metrics).

   Using the important property of any scalar fields minimally coupled to gravity,
\beq
	    T_0^0 = T_2^2 = T_3^3,                                     \label{TTT}
\eeq
  and the expressions \rf{Ric}, it is easy to find that \eqs \rf{EE'} combine to give
\beq
    \beta''=\gamma''= \mu'' = \thd\alpha''                     \label{bgm''}
\eeq
  (where the last equality holds due to \rf{harm}), whence it follows
\bear
		\mu \eql \thd(\alpha-Ax),
\nn
	     \gamma \eql \thd(\alpha-Bx),
\nn                                                              \label{bgm}
	      \beta \eql \thd(\alpha+Ax+Bx),
\ear
  where $A$ and $B$ are integration constants; other two constants are ruled 
  out by properly choosing the origin of the $x$ coordinate and the scale along the $z$ axis.

  The remaining equations for the unknowns $\alpha(x)$ and $\psi(x)$ read
\bearr						\label{a''}
		2\alpha'' = - 3 U \e^{2\alpha},
\yyy                               \label{psi''}
		12 n \psi'' = \e^{2\alpha} dU/d\psi,
\yyy						\label{int}
		\alpha'{}^2 - 9 n \psi'{}^2 = - \tfrac 32 U\e^{2\alpha} + \thd(A^2\!+\!AB\!+\!B^2),
\ear
   where \eqn{int} follows from \rf{G11} and is a first integral of \rf{a''} and \rf{psi''}.

  A value of $x$ where $\beta\to -\infty$, so that coordinate circles shrink to a point, 
  corresponds to an axis, whereas spatial infinity corresponds to $\beta\to\infty$.
  In the coordinates (\ref{harm}) the conditions \rf{ax1}, \rf{ax2} or those of a regular 
  asymptotic can only hold at $x\to\pm\infty$. Since $\mu$ and $\gamma$ must there
  tend to finite limits simultaneously, from \rf{bgm} it follows that a regular axis
  (say, at $x\to-\infty$) or a regular asymptotic (at $x=+\infty$) can exist if the 
  integration constants satisfy the requirement
\beq
			A = B = N >0.                                        \label{ABN}
\eeq
  Thus a regular axis and a regular asymptotic require the same relation \rf{ABN},
  which is favorable for the existence of solitonic solutions.

  Suppose there is a solitonic solution, regular at $x \to \pm \infty$. Then at both ends
  we have in the leading order of magnitude in $x$
\beq 					                        \label{abN}
             \alpha(x) \approx \beta(x)  \approx Nx.
\eeq
 
  At infinity, $x\to\infty$, choosing the $t$- and $z$-scales so that $\gamma = \mu \to 0$
  we can write 
\beq
       \alpha(x) = \beta(x) + o(1) \approx Nx ,                     \label{abNx}
\eeq
   hence the constant $N$ has a clear geometric meaning: according to (\ref{defect}),
\beq
	N = 1-\mu_s,                                           \label{Nmu}
\eeq
  where $2\pi \mu_s$ is the angular deficit at a string asymptotic.

  On the regular axis $x \to -\infty$, since the $t$- and $z$-scales 
  have already been fixed by the conditions at infinity, 
  we can only assume $\gamma = \mu \to \gamma\ax = \const$,
  hence, recalling the coordinate condition \rf{harm}, the regularity condition \rf{ax2},
  $\e^{\beta-\alpha}\beta' \to 1$, leads to
\beq  									\label{g_ax}
	    \e^{- 2\gamma\ax} N =1 \ \then\ \e^{2\gamma\ax} = N = 1-\mu_s.
\eeq
  It means that the value of the ``redshift function'' $g_{tt}= \e^{2\gamma}$ on the axis is
  directly related to the angular deficit at infinity.

  There is one more point of importance to bear in mind: in accord with \rf{ax1} and to 
  provide finiteness of $T\mN$ on the axis, it is necessary to require   
\beq                        \label{ax3}
		|\alpha''| \e^{-2\alpha} < \infty \ \ \ {\rm as} \ \ \  x\to -\infty,
\eeq
  from which it directly follows $|U| < \infty$ by \rf{a''}. At large positive $x$ no additional
  requirements are needed, we already have there $U\to 0$ due to \rf{abNx}. 

  As $x \to \infty$, we must also have $\e^{-2\alpha} \psi'{}^2 \to 0$ as a necessary condition of 
  vanishing curvature. The behavior of the scalar field $\psi$ is closely related to that of $\alpha$ 
  since, excluding $U$ from \rf{int} and \rf{a''} and using \rf{ABN}, we obtain
\beq  						\label{psi'}
		 9 n \psi'{}^2 = \alpha'{}^2 - \alpha'' - N^2.
\eeq
  In particular, on the axis $x\to - \infty$ we can assume $\alpha' = N + O(r) = N + O(\e^{N})$,
  hence due to \rf{psi'} we have  $\psi' = O(\e^{-N|x|/2})$ leading to a finite value of $\psi(-\infty)$.
  On the other hand, the requirement of a sufficiently fast decay of $\e^{-2\alpha}\psi'{}^2$ as
  $x \to \infty$ (see the end of Section 3) leads to a finite value of $\psi(+\infty)$. Thus the 
  $\psi$ field should vary in a finite range over all $x \in \R$. 

  Furthermore, integrating \eqn{a''} over all $x$, as long as $\alpha'(\pm\infty) = N$,
  we obtain
\beq
		\int_{-\infty}^{+\infty} U(\psi (x)) \e^{2\alpha}dx =0,      \label{int-U}
\eeq
  which means that $U(\psi) \ne 0$ has a variable sign in any solitonic solution.  

  Another important observation is that due to $A=B$ we have $\mu \equiv \gamma$,
  that is, $\og_{zz} = \og_{tt}$, hence {\it any solitonic solution in \ME\ is boost-invariant
  in the $(z,t)$ subspace.}

\subsection{Solitons in the Jordan frame}

  Let us find out how the above observations are modified if we seek a solitonic 
  solution in the Jordan frame \MJ. For its description, we can use the same functions
  $\psi, \alpha, \beta, \gamma, \mu$ as in \ME, obeying the same equations, but
  the boundary conditions should now be formulated for the metric \rf{ds_J}.

  We can write for any STT from the class \rf{S-STT}
\beq
		ds_J^2 = \e^{2\eta(x)} ds_E^2,
\eeq
  the function $\eta(x)$ depending on the choice of a theory. Therefore we have
  for the metrics \rf{ds_J} and \rf{ds_E} $\aJ = \alpha + \eta$ and similarly for other 
  metric coefficients. Consider the regular asymptotic requirements in \MJ, then 
  at the corresponding value $x = x_{\infty}$ we should have finite values of 
\[
      \gJ = \gamma + \eta \quad {\rm and} \quad \mJ = \mu + \eta
\]
  and an infinite value of $\bJ = \beta + \eta$. Since \eqs \rf{bgm} hold as before, 
  the above requirements can be satisfied only if $x_{\infty} = \pm \infty$ (let it be
  $+\infty$ without loss of generality), and $\mu - \gamma$ should be finite in this limit.
  From \rf{bgm} it then follows that $A = B = N > 0$ and $\mu \equiv \gamma$. We thus 
  come to a similar conclusion to the one made for solitonic solutions in \ME, now that
  $g_{tt} = -g_{zz}$. Even more than that: the same could be concluded if we required 
  a regular axis instead of a regular asymptotic (where, for consistency, we will put 
  $x = x\ax = -\infty$). We can state the following:

{\it  In any STT from the class \rf{S-STT}, a static vacuum \cyl\ solution with a regular 
  asymptotic and/or regular axis is necessarily boost-invariant in the $(z,t)$ subspace.}

  Thus it is unnecessary to postulate this stringy property \cite{vilenkin81} since it
  directly follows from the proper boundary conditions.

  Next, we can compare two expressions for $\beta$: one obtained from
  \rf{bgm}, with $A \!=\! B \!=\! N > 0$, that is, $\beta = \thd (\alpha \!+\! 2Nx)$,
  and the other, following from \rf{harm}: 
  $\beta  = \alpha - \gamma - \mu = \alpha - \gJ - \mJ + 2\eta$. Since $\gJ = \mJ$
  tend to finite limits as $x \to \infty$, this comparison gives in the same limit:
\beq                    \label{ab+}
          \alpha = Nx - 3\eta + O(1), \quad\    \beta = Nx - \eta + O(1).     
\eeq
  Let us now apply the boundary condition \rf{defect} in \MJ: as $x\to \infty$,
\beq        \label{circ+}
		\e^{\bJ-\aJ} \bJ' \sim \e^{2\eta} (\beta' + \eta') \sim \e^{2\eta} N = O(1),
\eeq
  whence it follows that $\eta(x)$ tends to a finite limit at large $x$. As a result, 
  we have finite limits of the Einstein-frame quantities $\gamma$ and $\mu$ and 
  also $\alpha \sim \beta \sim Nx$, in full similarity with the Einstein frame. 
  In \rf{circ+} we have used the fact that $\eta' = (d\eta/d\psi) \psi'$ vanishes as 
  $x\to\infty$ since $\psi' \to 0$ (see \rf{psi'}) while $d\eta/d\psi$ must be finite 
  due to the the regularity requirement for $\eta(\psi)$. 

  Quite a similar reasoning can be used for $x\to - \infty$ (the axis), with the 
  difference that now it is required $\e^{\bJ-\aJ} \bJ' \to 1$. 
  
  So far, in this subsection, we did not make any assumptions on the properties 
  of a solution in \ME\ and even did not use the evident requirement that $\eta(\psi)$ 
  should be regular in the whole range $x \in \R$, considering each limit $x\to \pm \infty$
  separately and using the transformation \rf{J-E} as simply a substitution in the field 
  equations. Invoking the boundary conditions, we have concluded that $\eta(\psi)$
  must not only be regular but also have finite limits on the axis and at infinity.  

  We can now assert that the the limit $x\to -\infty$ we have finite values of $\eta=\eta\ax$,
  $\mu = \gamma = \gamma\ax$ as well as $\bJ - \aJ = \beta-\alpha$; in addition, 
  $\eta' \to 0$, therefore, in this limit, 
\beq        \label{circ-}
		\e^{\bJ-\aJ} \bJ' = \e^{\beta-\alpha} \beta' = 1,
\eeq
  as required in \rf{ax2}. Thus {\it a regular axis in \MJ\ implies a regular axis in \ME.}

  At the asymptotic $x\to +\infty$, the first equality of \rf{circ-} again holds, which means 
  that {\it a regular asymptotic in \MJ\ implies a regular axis in \ME.} Even more than that,
  the angular defects in \ME\ and \MJ\ coincide, $\mu_s = 1 - N$. What is different, is 
  the relationship between $N$ and the ``redshift function'' $\e^{2\gJ}$: since 
  $\gJ = \gamma + \eta$, we now have  
\beq
  			\e^{2\gJ}\Big|_{x\to -\infty} = N \e^{2\eta\ax}.
\eeq

  Thus we have the following general result:

{\it If the conformal factor $\e^{2\eta}$ is finite in the whole range $x\in \R$ and 
   in the limits $x\to \pm\infty$, then solitonic solutions exist simultaneously in \MJ\ and
   in \ME\ and are characterized by the same deficit angle.} 

  A possible singular behavior of the conformal factor violates this correspondence,
  as is evident from the example \rf{cos}: if we have a solitonic solution
  in \ME\ but $\cos\psi = 0$ at some $x$, on such a surface we have a singularity 
  in \MJ. It is an evident case of a conformal continuation, in which the transformation
  \rf{J-E} maps the whole singular manifold \MJ\ to only a part of the regular
  manifold \ME.

  We also cannot exclude a contrary situation, that there is a solitonic solution in \MJ\
  but this manifold (or its part) maps to a singular manifold \ME, similarly to what
  is described, for example, in \cite{kb73}, where \MJ\ is a \wh\ with a conformal scalar 
  but only its region maps to a singular (Fisher) space-time \ME.

  The important constraint \rf{int-U} on the scalar field potential obtained in \ME\ 
  easily transforms to \MJ: indeed, according to \rf{J-E}, we find that 
  $U(\psi) = \e^{4\eta}V(\phi)$ and $\e^{2\alpha} = \e^{2\aJ - 2\eta}$ and rewrite 
  \eqn{int-U} in terms of the quantities specified in \MJ\ as
\beq                    \label{int-V}
		\int_{-\infty}^{+\infty} V(\phi) \e^{2\aJ + 2\eta} dx =0,
\eeq
  where $x$ is, as before, a harmonic coordinate in \ME. 

  The whole content of this subsection applies to any STT from the class \rf{S-STT}.

 \section{Examples}

  The first step in attempts to find solitons in \MJ\ as described in the previous section 
  is to solve the set of equations 
\bearr						\label{a''1}
		2\alpha'' = - 3 U \e^{2\alpha},
\yyy                               \label{psi''1}
		12 n \psi'' = \e^{2\alpha} dU/d\psi,
\yyy						\label{int1}
		\alpha'{}^2 - 9 n \psi'{}^2 = N^2 - \tfrac 32 U\e^{2\alpha},
\ear
  with the unknowns $\alpha(x)$ and $\psi(x)$, constants $n = \pm 1$ and $N > 0$,
  while the choice of the potential $U(x)$ corresponds to the choice of a particular HMPG 
  theory. A combination of \rf{a''1} and \rf{int1} free from $U$ is
\beq                        \label{psi'1}
		9 n \psi'{}^2 = \alpha'{}^2 - \alpha'' - N^2 .
\eeq

  There are some examples of $U(x)$ with which \eqs \rf{a''1}--\rf{int1} can be 
  solved analytically, but in general, with given $U(\psi)$, numerical methods are 
  necessary. An alternative way is to use inverse problem methods \cite{kb01-cy},
  of which the simplest is to specify the function $\alpha(x)$ or $\alpha'(x)$ and to 
  find $\psi(x)$ and $U(x)$ from the equations; the function $U(\psi)$ is then well 
  defined if $\psi(x)$ is monotonic. An advantage of this method is that from the 
  previous analysis we know much about the behavior of $\alpha(x)$ suitable for 
  a solitonic solution. 

  The following algorithm can be suggested for finding a solitonic solution in \MJ:
\begin{enumerate}
\item
	Specify $\alpha(x)$ regular in $x\in \R$ and such that $\alpha'(\pm \infty) = N >0$.
\item 
	Verify the suitable behavior of $\alpha(x)$: that\\  
     (i) $|\alpha''| \e^{-2\alpha} \!< \! \infty$ as $x\! \to\! -\infty$ (see \rf{ax3}) and 
     (ii) that the r.h.s. of \eqn{psi'1} has a definite sign at all $x \in \R$ 
     (we obtain $n\,{=}\,1$, the canonical sector, if it is nonnegative, and $n\,{=}\,-1$ otherwise). 
\item 
	Find $\psi(x)$ from \eqn {psi'1}. 
\item
	Verify that the range of $\psi$ is within the validity range of the map \rf{J-E}.
\item
	Find $U(x)$ from \eqn{a''1}.
\item
	With the solution found, verify the validity of \eqn{psi''1}.
\end{enumerate}

   Step 2(ii) is necessary as long as we adhere to systems with either a canonical or a phantom 
   scalar field, excluding transitions from one sector to another within a single solution 
   (such solutions do exist and have been considered as possible sources for
   \whs\ with so-called ``trapped ghosts'' \cite{trap1, trap2, kb18-wh}, this name used 
   because a phantom field exists there is a bounded part of space).  
   
   For step 4, it is important that only derivatives of $\psi$ appear in \eqs \rf{a''1}--\rf{psi'1},
   therefore one can add to $\psi$ an arbitrary constant to adjust its range to that of the map 
   \rf{J-E}.
  
  The last step is needed only for confidence since \rf{psi''1} is a consequence of 
  \rf{a''1} and \rf{int1}.

\subsubsection*{Example 1: Zero potential}

  This example is trivial, it does not promise any solitonic configurations, and we present 
  it here only for completeness and clarity. The corresponding solution in \ME\ is well known, 
  it is a direct extension of the famous century-old Levi-Civita solution \cite{Levi,BSW19}. 
  In our coordinates and notations, the whole set of equations in \ME\ reduces to
\bearr            \label{eq-0}
         \psi'' = \beta'' = \gamma''= \mu'' = 0,
\nnn
	    \beta'\gamma'  + \beta'\mu' + \gamma' \mu' = 3n \psi'{}^2 ,
\ear
  and its solution can be written, without loss of generality, as
\bearr  \nhq                    \label{ds0}
		ds_E^2 =\! \e^{2cx} dt^2\! -\! \e^{2ax}dx^2\! -\! \e^{2mx}dz^2\! - \! \e^{2bx}d\varphi^2,
\yyy					\label{int0}
		\psi = \psi_0 + C x, \quad\  bc+ bm + cm = 3n C^2,
\ear
  where $b,c,m,\psi_0,\psi_1$ are integration constants, and $a=b+c+m$. The requirement
  of flatness at large $x$ leads to $c= m =0$, hence by \rf{int0} we have $\psi_1=0$
  (the scalar field is trivial) and a Minkowski metric up to an angular deficit $\mu_s = 1-b$.
  The metric is globally flat if $b=1$ and has a conical singularity on the axis ($x\to -\infty$)
  if $b \ne 1$. 

\subsubsection*{Example 2: Exponential potential}

   Assuming the potential in the form 
\beq  				\label{U2}
  			U(\psi) = U_0 \e^{\lambda \psi}, \quad U_0, \lambda = \const,
\eeq
   it is easy to solve \eqs \rf{a''1}--\rf{int1} analytically, but it is impossible 
   to obtain a stringlike solution because of a constant sign of $U$, recall \rf{int-U}.

  Indeed, with \rf{U2}, combining \eqs \rf{a''1} and \rf{psi''1}, we get
\beq                          \label{psi'2}
               \lambda \alpha'' + 18 n\psi''=0 \ \then \     \lambda \alpha' + 18 n\psi'=  C,  
\eeq
  where the integration constan $C$ can be fixed from the regular axis conditions:
  since, as we know, $\alpha' \to N$ and $\psi' \to 0$ as $x \to -\infty$, we have 
  $C = \lambda N$. 

  Substituting \rf{psi'2} with $C = \lambda N$ to \rf{int1}, we obtain
\beq               \label{a'2}
          \alpha'{}^2 - \frac {n\lambda^2}{36} (N-\alpha')^2 - N^2 
                                                     = \frac 32 U_0\e^{\lambda\psi+2\alpha}.        
\eeq
  At a regular asymptotic $x\to\infty$ we must have $\alpha' \to N$, so that the l.h.s 
  of  \eqn {a'2} turns to zero, whereas the r.h.s. tends to infinity due to 
  $\alpha\approx Nx$ while $\psi$ should tend to a finite constant. This contradiction 
  shows that assuming a regular axis, it is impossible to obtain a regular asymptotic. 

  Therefore we will stop here, even though \eqn{a'2} can be easily further integrated 
  (see, e.g., \cite{kb01-cy} for a solution in slightly different notations). Let us only 
  note that our observations on the exponential potential include the special case of 
  a constant potential, $\lambda =0$ in \rf{U2}, equivalent to a cosmological constant 
  in \ME.  

\subsubsection*{Example 3: Solitons with nonzero potential on the axis}
\def\ka{(k\alpha)}

  The calculations turn out to be easier if we specify $\alpha'$ as a function of $\alpha$
  instead of $\alpha(x)$. The following example uses a suggestion from \cite{kb01-cy},
  where it was discussed in the framework of GR: we suppose
\bearr                           \label{ex3}
              \alpha'{}^2 = N^2 \bigg(1 + \frac H {\cosh^2 (k\alpha)}\bigg), 
\nnn
		k = \const >0, \quad  H = \const > -1.
\ear
   From \rf{ex3}, with \rf{psi'1}, it is easy to obtain
\bearr                          \label{a''3}
  		    \frac d{d\alpha} (\alpha'{}^2) = 2 \alpha'' = - \frac{2HN^2 k \sinh\ka}{\cosh^3 \ka},
\yyy                            \label{psi'3}
		   9n\psi'{}^2 = \frac{H N^2}{\cosh^2 \ka}\Big(1 + k \tanh \ka\Big) ,
\ear
  where, as before, the prime denotes $d/dx$. 

  From \rf{psi'3} it follows that to keep the same sign of $n\psi'{}^2$ for all $\alpha$
  (and $x$ as well) we should require $k \leq 1$. On the other hand, on the axis $x\to -\infty$, 
  we have $\alpha \sim Nx \to -\infty$, and by \rf{psi'3} $\alpha'' \sim \e^{-2k |\alpha|}$,
  therefore, to have a finite limit of $\alpha'' \e^{-2\alpha}$ according to \rf{ax3} 
  it is necessary to require $k \geq 1$. Thus we arrive at the unambiguous value $k = 1$.

  Putting $k=1$, by direct calculation of $x = \int d\alpha/{\alpha'}$ with \rf{ex3},
  we obtain without loss of generality
\beq 			\label{alpha-x}
               \sinh (Nx) = \frac{\sinh \alpha}{\sqrt{1+H}},     
\eeq
  thus, as required, $\alpha\in \R$, and its infinities coincide with those of $x$.
  The potential $U(x)$ is determined by \rf{a''1} and hence by \rf{a''3}.
  
  We deal with the canonical sector if $H>0$ and with the phantom sector if $H < 0$.  
  For calculations, it is more convenient to pass on in \rf{psi'3} from $\psi'$ to 
  $\psi_\alpha \equiv d\psi/d\alpha= \psi'/\alpha'$:
\beq		
		9n\psi_\alpha^2 = 9n\frac {\psi'{}^2}{\alpha'{}^2}=
					\frac{H(1+\tanh\alpha)}{H+\cosh^2\alpha}.
\eeq
  In all cases we see that at large $|\alpha|$ the expression for $\psi'$ vanishes exponentially, 
  hence $\psi$ has finite limits $\psi(\pm\infty) = \psi_{\pm}$. Thus in the canonical sector, where 
  the conformal factor $\e^{2\eta} = \cosh^2\psi$ is specified for $\psi\in R$, we certainly obtain 
  a family of solitonic solutions in \MJ, while in the phantom sector ($\e^{2\eta} = \cos^2\psi$) 
  we must select such parameter values that the range of $\psi$ is located within a single
  half-wave of $\cos\psi$. Let us discuss them separately and graphically illustrate the 
  behavior of solitonic solutions for selected values of the parameters.

\paragraph{The canonical sector, $n=1$.} 
  Let us put for certainty $H=1/2$. Using \rf{alpha-x} and \rf{psi'1}, the function  
  $\psi(x)$ for given values of $N$ is determined by the integral
\beq                   \label{psi-int}
		  \psi(x)=\int_0^x \sqrt{\frac{\alpha'(\xi)^2-\alpha''(\xi)-N^2}{9n}}d\xi.
\eeq
  and is plotted in Fig.\,1 for selected values of $N$. Furthermore, using the expressions
\beq                  \label{Ur}
		U=-\frac{2}{3}\alpha''\e^{-2\alpha}, \quad r=\exp\left(\frac{\alpha + 2N x}{3}\right)
\eeq
 and the relation \rf{alpha-x}, we plot the potential $U$ as a function of the circular radius 
 $r = \e^\beta$ in Fig.\,2.
\begin{figure}[t]
\centering
\includegraphics[width=0.95\linewidth]{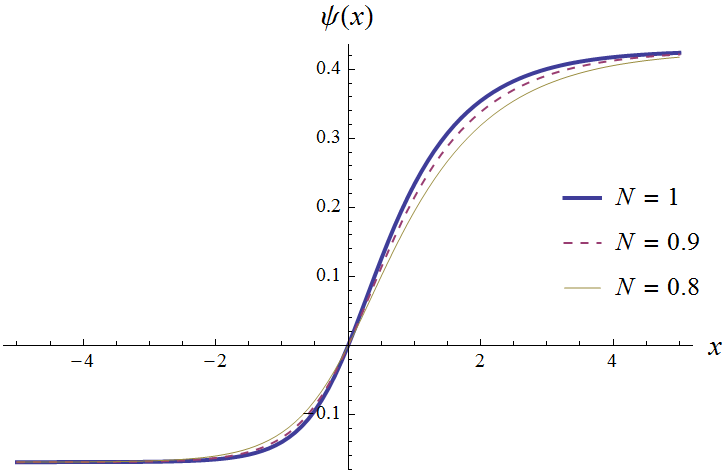}
\caption{\small
	Plots of $\psi(x)$ for $H=1/2$ (the canonical sector) and $N=1, 0.9, 0.8$.}
\label{ex3can-psix}
\end{figure}
\begin{figure}[t]
\centering
\includegraphics[width=0.95\linewidth]{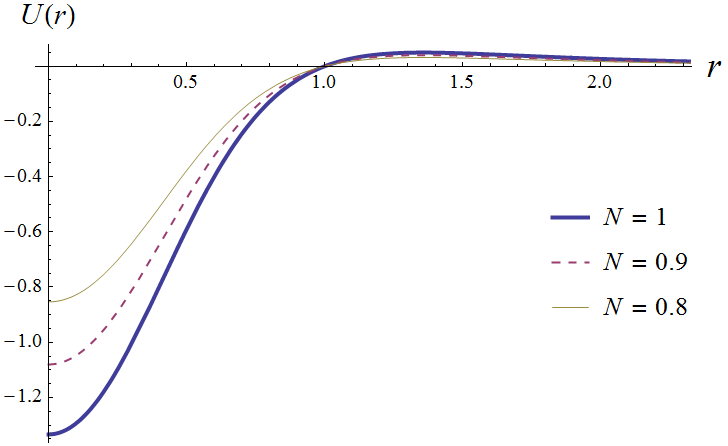}
\caption{\small
	The potential $U(r)$ in the Einstein frame for $H=1/2$ and $N=1, 0.9, 0.8$.}
\label{ex3can-U}
\end{figure}
\begin{figure}[h!]
\centering
\includegraphics[width=0.95\linewidth]{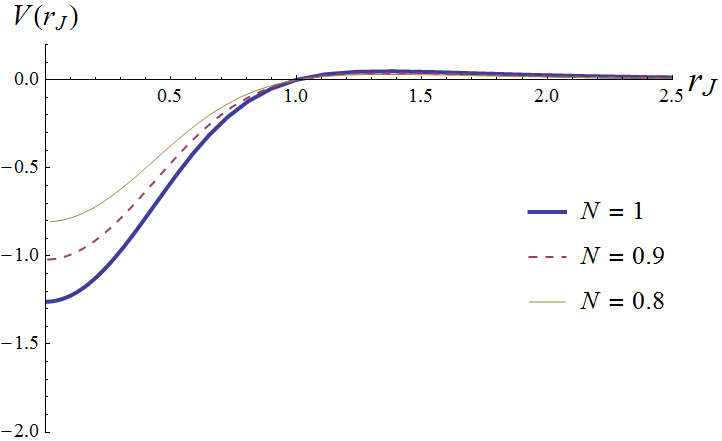}
\caption{\small
	The potential $V(\rJ)$ in the Jordan frame for $H=1/2$ and $N=1, 0.9, 0.8$.}
\label{ex3can-V}
\end{figure}

  To obtain a similar picture for the potential $V$ in \MJ, we use relations between the 
  potentials and the circular radii $r$ and $\rJ = \e^\bJ$ in the two frames:
\bearr 
		V=(1+\phi)^2U= -\frac{2}{3} \cosh^{-4}\psi \, \alpha''\e^{-2\alpha},
\nnn
		\rJ = r\cosh\psi = \exp\left(\frac{\alpha + 2N x}{3}\right)\cosh\psi.
\ear	
  The results for the same parameter values as in Figs.\,1 and 2 can be seen in Fig.\,3
  and are qualitatively similar to those in Fig.\,2.

\paragraph{The phantom sector, $n=-1$.} 
  In this case we can put $H=-1/2$. Using the same methodology as for the 
  canonical sector, we plot $\psi(x)$ obtained from \rf{psi-int} and \rf{alpha-x} in Fig.\,4.
    
  The range of $\psi(x)$, coinciding with the difference between its asymptotic values,
\beq
  \Delta\psi_{\pm}=\psi(+\infty)-  \psi(-\infty)\approx 0.795 < \pi,
\eeq
   is located within a single half-wave of $\cos\psi$, as required.
  
  The potential $U(r)$ in the Einstein frame is obtained in the same way as for the 
  canonical sector and is plotted in Fig.\,5. To obtain the potential $V(\rJ)$ in \MJ,
  we use the following expressions:
\bearr 
   		V=(1+\phi)^2U= -\frac{2}{3\cos^4\psi}\,\alpha''\e^{-2\alpha},
\nnn
  		r_{\rm J}=r|\cos\psi|=\exp\left(\frac{\alpha + 2N x}{3}\right)|\cos\psi|.
\ear	
  Figure 6 shows the corresponding parametric plots.
  
  It is easy to notice that all potentials in the stringlike solutions have a changing sign,
  in agreement with \rf{int-U} and \rf{int-V}.
  Also, in the canonical sector, the potential close to the axis has an attracting nature.
  In the phantom sector it looks repulsive, but it is still attractive for a phantom field
  which in general tends to climb a potential instead of rolling down. 
     
\begin{figure}[h]
\centering
\includegraphics[width=0.95\linewidth]{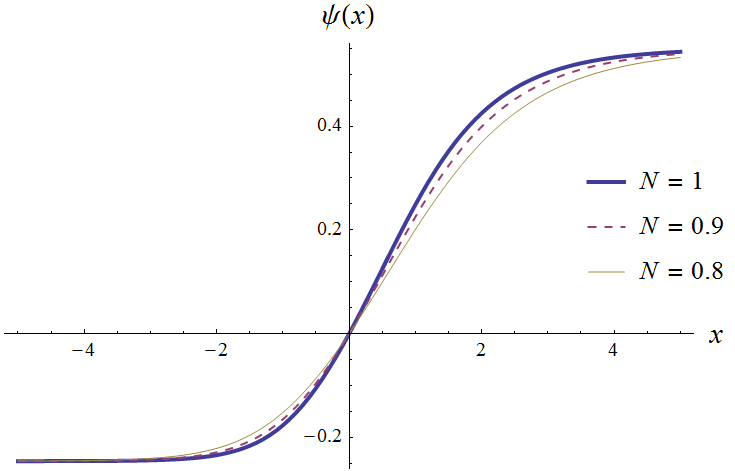}
\caption{\small
	The scalar field $\psi(x)$ for $H=-1/2$ (the phantom sector) and $N=1, 0.9, 0.8$.}
\label{ex3phant-psix}
\end{figure} 
\begin{figure}[h]
\centering
\includegraphics[width=0.95\linewidth]{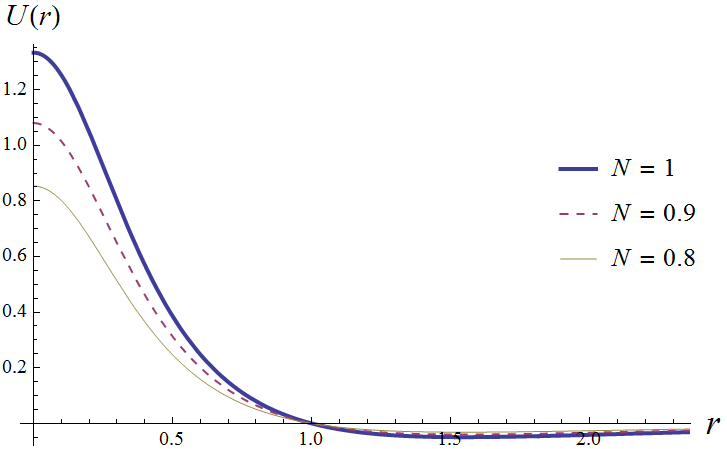}
\caption{\small
	The potential $U(r)$ in the Einstein frame for $H=-1/2$ (phantom sector) and $N=1, 0.9, 0.8$.}
\label{ex3phant-U}
\end{figure}  
\begin{figure}[h!]
\centering
\includegraphics[width=0.95\linewidth]{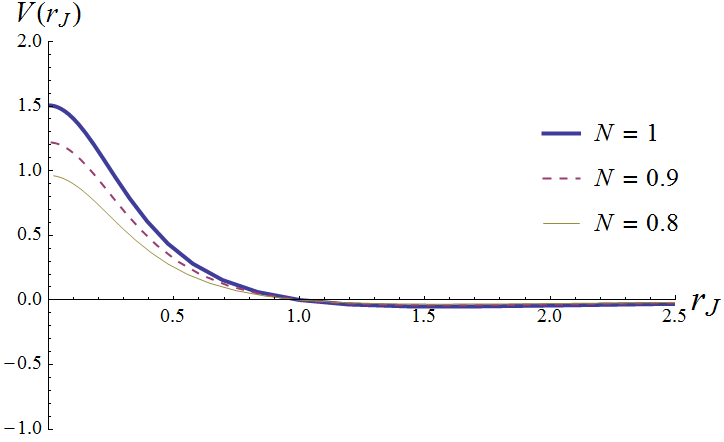}
\caption{\small
	The potential $V(r_{\rm J})$ in the Jordan frame for $H=-1/2$ (phantom sector) and $N=1, 0.9, 0.8$.}
\label{ex3phant-V}
\end{figure}

\section{Concluding remarks} 

  We have considered the opportunity of obtaining stringlike, or solitonic vacuum solutions with 
  cylindrical symmetry in the framework of hybrid metric-Palatini gravity (HMPG) without specifying 
  the dependence $f(\cR)$, or, equivalently, the potential $V(\phi)$ in the STT representation of 
  this theory. In addition to producing some particular examples of such configurations, we 
  have obtained some results of more general significance:

\begin{enumerate}
\item
     Equations \rf{int-U} and \rf{int-V} implying that stringlike solutions cannot be obtained with 
     purely nonpositive or purely nonnegative potentials.
\item
     Vacuum static \cyl\ solutions with a regular asymptotic or a regular axis are necessarily 
     boost-invariant in the $(z,t)$ subspace.
\item
   If the conformal factor $\e^{2\eta}$ between \MJ\ and \ME\ is regular in the whole range 
   $x\in \R$ and finite in the limits $x\to \pm\infty$, then solitonic solutions exist 
   simultaneously in \MJ\ and in \ME\ and are characterized by the same deficit angle.
\end{enumerate}

  All these observations are not restricted to the STT representation of HMPG but apply
  to an arbitrary STT from the class \rf{S-STT}; moreover, they apply to arbitrary $f(R)$
  theories of gravity since the latter are known to coincide with the Brans-Dicke STT
  with the coupling constant $\omega =0$ and a nonzero scalar field potential whose form
  depends on the function $f(R)$ \cite{ex-GR1, ex-GR2}.

  We have discussed here only static \cyl\ configurations, but probably of even greater 
  interest can be stringlike models with rotation, for which numerous solutions are known
  in GR (see, e.g., \cite{exact-book, BSW19, we19}), and their extensions to HMPG
  are quite possible.

  In addition to stringlike solitons with a regular axis, possible objects of interest,
  which can look from outside like cosmic strings, are \cy\ \whs\ which are also 
  globally regular configurations but which do not contain a 
  symmetry axis at all, similarly to spherically symmetric \whs\ which do not contain 
  a center of symmetry. Instead of an axis, where the circular radius $r$ turns to zero, 
  \cy\ \whs\ have a minimum of $r$ and, around it, two regions with much larger values 
  of $r$ \cite{kb-lem09, kb-lem13}. 

  It is well known that the existence of \whs\ in GR in general requires some 
  amount of ``exotic'' matter violating the standard (Weak and Null) energy conditions,
  and its necessity can be avoided either in alternative theories of gravity 
  (see \cite{viss-book, lobo17-ed} for reviews) or within \GR\ in 
  \cy\ symmetry by invoking rotation and an appropriate choice of matter sources 
  \cite{kb-kr19, we19-wh, kb20-wh}. In this respect we can note that HMPG,
  being a purely geometric source of scalar fields, both canonical and phantom, 
  can provide a natural framework for \wh\ construction without other ``exotic'' sources. 
  Within spherical symmetry this has already been confirmed \cite{kb19, we20}, and a 
  similar consideration in \cy\ symmetry can be one of the subjects of our further studies. 
    
 \bigskip 
    
\subsection*{Acknowledgments}

The work was funded by the RUDN University Program 5-100 and the Russian 
 Basic Research Foundation grant 19-02-0346.
The work of K.B. was also partly performed within the framework of the Center FRPP supported 
by MEPhI Academic Excellence Project (contract No. 02.a03.21.0005, 27.08.2013).

\bigskip

\end{document}